\documentclass[a4,aps,twocolumn,showpacs,floatfix,longbibliography]{revtex4-1}
\usepackage{graphicx}
\usepackage{dcolumn}
\usepackage{color}
\usepackage{amsmath}
\usepackage{here}
\usepackage{systeme}
\usepackage{xcolor}
\usepackage{multirow}

\RequirePackage{natbib}

\begin{document}

\title{Acoustic analog of Hawking radiation in quantized circular superflows
of Bose-Einstein condensates}
\author{ Igor Yatsuta$^1$, Boris Malomed$^{2,3}$, Alexander Yakimenko$^1$}
\affiliation{$^1$ Department of Physics, Taras Shevchenko National University of Kyiv,
64/13, Volodymyrska Street, Kyiv 01601, Ukraine \\
$^2$ Department of Physical Electronics, School of Electrical Engineering,
Faculty of Engineering, and Center for Light-Matter Interaction, Tel Aviv
University, 69978 Tel Aviv, Israel \\
$^3$ Instituto de Alta Investigaci\'{o}n, Universidad de Tarapac\'{a},
Casilla 7D, Arica, Chile}

\begin{abstract}
We propose emulation of Hawking radiation (HR) by means of acoustic
excitations propagating on top of persistent current in an atomic
Bose-Einstein condensate (BEC) loaded in an annular confining potential.
The setting is initially created as a spatially uniform one, and then
switches into a nonuniform configuration, while maintaining uniform BEC
density. The eventual setting admits the realization of sonic black and white
event horizons with different slopes of the local sound speed. A smooth
slope near the white-hole horizon suppresses instabilities in the supersonic
region. It is found that tongue-shaped patterns of the density-density
correlation function, which represent the acoustic analog of HR, are
strongly affected by the radius of the ring-shaped configuration and number
of discrete acoustic modes admitted by it. There is a minimum radius that
enables the emulation of HR. We also briefly discuss a possible similarity
of properties of the matter-wave sonic black holes to the known puzzle of
the stability of Planck-scale primordial black holes in quantum gravity.
\end{abstract}

\maketitle

\section{Introduction}

\label{sec:Intro} Black holes (BHs) are among the most fascinating objects
in the Universe, the study of which may help to illuminate an intimate
relation between gravity and the quantum theory. Pioneering works \cite%
{bekenstein1973, hawking1974, hawking1975,hawking1976} of Bekenstein and
Hawking had allowed elucidating the remarkable quantum properties of BHs. It
was discovered that BH is not a stationary everlasting object, as it emits
black-body radiation.
The effective temperature of the Hawking radiation (HR) is extremely low for
known astrophysical BHs, which makes it practically impossible to observe
the emission effect,
therefore attention has turned to settings that admit emulation of this
phenomenology in other physical settings.

Acoustic BHs were first introduced by Unruh \cite{unruh1976,unruh1981}, who
had demonstrated precise formal equivalence between the behavior of sound
waves in fluid flow and that of a scalar field in curved space-time. Thus,
it is possible to create analog BH in superfluids where the transition from
subsonic to supersonic flow plays the role of the event horizon. Following
that work, many experimental realizations were proposed for demonstrating
acoustic horizons. Behaviors similar to the analog HR have been
experimentally and theoretically explored in trapped ions \cite%
{horstmann2010}, optical fibers \cite{philbin2008,belgiorno2010, unruh2012,
liberati2012}, electromagnetic waveguides \cite{schutzhold2005}, water tanks
\cite{weinfurtner2011,rousseaux2008}, ultracold fermions \cite%
{giovanazzi2005}, exciton-polariton condensates \cite{solnyshkov2011} and
superfluid $^{3}$He \cite{jacobson1998}. Tremendous progress in physics of
ultracold atomic gases \cite{pitaevskii2016} makes atomic Bose-Einstein
condensates (BECs) very suitable testbeds for sonic BHs \cite{carusotto2008,
mayoral2011}. 
In these connections, recent experimental realization \cite%
{lahav2010,steinhauer2016,steinhauer2019,aless2020rampup} of analogs of the
HR in atomic BEC is an important milestone in the emulation of quantum
properties of BHs.

Most previous works addressing sonic BHs in BECs were focused on
cigar-shaped quasi-one-dimensional (1D) condensates \cite{steinhauer2016},
or ones subject to specific absorbing boundary conditions used for
suppression of spurious instabilities \cite{carusotto2008,mayoral2011},
which give rise to a single BH horizon. 
On the other hand, toroidal BECs were used to emulate
quantum features of HR, as the angular momentum is quantized in the
ring-shaped BEC due to periodic boundary conditions. In this work we address
quasi-1D atomic BEC confined in a toroidal trap. This setting, which is
available in experiments \cite{24b,24c,24d,24e,B_rkle_2018}, was used for
studies of diverse matter-wave patterns maintained by circular flows, i.e.,
effects induced by the angular momentum \cite%
{24f,24g,24h,24i,24j,24k,24l,24m,24n,24o,24p,24q,24r,Bland20,walczak2011}.


Previous investigations \cite{garay2001, jain2007} of the event horizon in
BEC with toroidal geometry relied upon the use of the supersonic flow driven
by a variable density distribution in the condensate.
A spectrum of the eigenstates on top of the nonuniform background was found
in Ref. \cite{jain2007}, demonstrating parametric amplification (i.e.,
dynamical instabilities) at certain frequencies, and it was compared to the
two-mode approximation \cite{garay2001,jain2007}. 
Significant theoretical progress in studies of acoustic analogs of HR was
achieved with the help of the correlation-function method, which was
extensively used in the context of elongated 1D condensates \cite%
{carusotto2008,mayoral2011,steinhauer2016,steinhauer2019}. However, not much
attention was paid to the white-hole (WH) horizon, and the role of the
spatial slope of the nonuniform local parameters of the condensates near
this point, on the stability of the system. 

\begin{figure}[th]
\centering
\includegraphics[width=.99\linewidth]{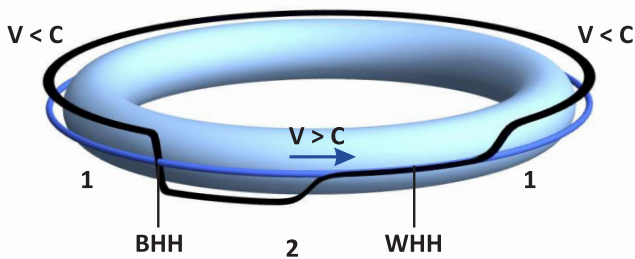}
\caption{A schematic of the ring-shaped Bose-Einstein condensate with
asymmetric acoustic event horizons. Shown are the blue isosurface of the
azimuthally uniform condensate density $n$, and the constant superflow
velocity, $v$ (the dark blue line). The azimuthally-modulated interaction
strength, $g(x,t)$, is responsible for the variation of the local sound
speed $c(x)$ (the black line), which identifies the subsonic (1: $v<c$) and supersonic (2: $%
v>c $) regions. The arrow shows the direction of the persistent current.
Acoustic black-hole and white-hole horizons (BHH and WHH, respectively) are
located at points with $c(x)=v$.}
\label{model}
\end{figure}

Here we consider the setting with a \emph{uniform condensate density}, while
the event horizon is created by spatiotemporal modulation of the coupling
constant, which is responsible for interatomic interactions in the
mean-field approximation \cite{pitaevskii2016,carusotto2008}. The
coexistence of BH and WH event horizons in the toroidal geometry gives rise
to amplified HR and formation of background ripples \cite{steinhauer2014,
steinhauer2017,corley1999, wang2017, wang20172}. Extensive investigation of
related instabilities be dint of the correlation-function method has been
carried out in the 1D system with absorbing boundary conditions \cite%
{carusotto2016}. Here we focus on the HR per se, rather than the
amplification effect induced by the instability. As is known \cite%
{unruh1981,1998CQGra..15.1767V,steinhauer2019}, the analog Hawking
temperature for the effectively 1D flow is
\begin{equation}
T_{H}=\frac{\hbar }{2\pi k_{B}}\left( \frac{\partial v}{\partial x}-\frac{%
\partial c}{\partial x}\right) \Big |_{x=x_{h}},  \label{eq:T_H}
\end{equation}%
where $v(x)$ is the local velocity of the condensate, $c(x)$ is the local
speed of sound, and the derivatives are evaluated at the position of the
horizon, $x=x_{h}$. To minimize the impact of the WH horizon, we introduce a
steep gradient of $c(x)$ near the BH horizon, and a smoothed gradient near
the WH one, as shown in Fig. \ref{model}. This setting makes it possible to
conveniently manipulate the slopes, $\partial v/\partial x$ and $\partial
c/\partial x$ in Eq. (\ref{eq:T_H}), while keeping parameters of the
acoustic BH horizon and density of the condensate unchanged. 

The main objective of the present work is the impact of the quantization of the
superflow, imposed by the periodic boundary conditions, on the analog HR in
the toroidal BEC. The analysis reveals the following noteworthy effects. (i)
A smooth WH horizon suppresses instabilities and makes the supersonic flow
robust. (ii) Density-density correlations, representing the acoustic analog
of HR, are strongly affected by the size of the ring's radius and the number of
discrete acoustic modes available in the annular geometry initially. (iii) HR vanishes
if the radius falls below a critical value.



The rest of the paper is organized as follows. The model is introduced in
Sec. \ref{sec:model}. In Sec. \ref{sec:fluctuations_and_dispersion} we
discuss initial fluctuations and the corresponding dispersion relation. In
Sec. \ref{sec:Hawking_radiation} we present results for density-density
correlations and the acoustic analog of HR near the BH horizon.
The paper is concluded by Sec. \ref{sec:conclusions}, where we also discuss
possible similarities of properties of the analog BHs to some fundamental
problems, such as stability of primordial quantum BHs on the Planck's scale.

\section{The model}

\label{sec:model}
We address the acoustic analog of HR in the framework of the 1D
Gross-Pitaevskii equation (GPE), written in scaled units,
\begin{equation}
i\frac{\partial \psi }{\partial t}=\left[ -\frac{1}{2}\frac{\partial ^{2}}{%
\partial x^{2}}+V(x,t)+g(x,t)|\psi |^{2}\right] \psi ,  \label{eq:GPE1D}
\end{equation}%
where $\psi (x+L,t)=\psi (x,t)$ is the wave function of the ring-shaped
condensate of ring $R$ and length $L=2\pi R$, while $V(x,t)$ and $g(x,t)$
are the effective potential and 1D coupling constant, respectively. An
efficient method for detecting the analog HR is the analysis of
density-density correlations 
\cite{balbinot2008, parentani2010, steinhauer2015}. We here use mean-field
simulations of the GPE, together with a sampling of the quantum noise by means
of a procedure similar to the well-known truncated Wigner approximation
(TWA) \cite{sinatra2001, sinatra2002, ruostekoski2013,
opanchuk2013,drummond2017}. In our analysis of the quantum fluid dynamics,
we consider the evolution of stochastic trajectories satisfying the 1D GPE (%
\ref{eq:GPE1D}). Similar to the standard TWA, we add extra noise to the
initial wave function, thus taking it as follows:
\begin{equation}
\begin{split}
\psi (x,0)& =e^{ik_{0}x}\Big[\sqrt{n_{0}}+\frac{1}{\sqrt{L}}\sum_{k\neq 0}%
\big(\beta _{k}u_{0k}e^{ikx}+ \\
& +\beta _{k}^{\ast }w_{0k}^{\ast }e^{-ikx}\big)\Big],
\end{split}
\label{eq:inNoise}
\end{equation}%
where $k_{0}=Mv/\hbar $ is determined by the flow velocity $v$, $\beta
_{k},\beta _{k}^{\ast }$ are noise amplitudes taking random values, the
summation is performed over wavenumbers $k$, $n_{0}$ is the spatially
uniform density, and $u_{0k},w_{0k}$ are defined as
\begin{equation}
u_{0k},w_{0k}=\big [(E_{k}/\epsilon _{k})^{1/2}\pm ((E_{k}/\epsilon
_{k})^{-1/2}\big]/2,  \label{eq:coefs}
\end{equation}%
where $E_{k}\equiv k^{2}/2$ , $\epsilon _{k}\equiv \sqrt{E_{k}(E_{k}+2gn)}$.

It is relevant to briefly discuss the derivation of GPE (\ref{eq:GPE1D}) and
the meaning of its parameters, as well as the meaning of the noise in Eq. (%
\ref{eq:inNoise}). As said above, we consider the system schematically shown
in Fig. \ref{model}. The azimuthally-modulated interaction strength, $g(x,t)$%
, where $x=R\varphi $ and $\varphi $ is the azimuthal coordinate along the
ring, is responsible for the variation of the local sound speed, $c(x)$. The
starting point of the derivation is the 3D mean-field GPE in physical units:
\begin{equation}
i\hbar \frac{\partial \bar{\Psi}}{\partial \bar{t}}=\left[ -\frac{\hbar ^{2}%
}{2M}\bar{\nabla}^{2}+\bar{V}_{e}(\bar{\mathbf{r}},\bar{t})+\gamma (\bar{%
\mathbf{r}},\bar{t})|\bar{\Psi}|^{2}\right] \bar{\Psi},  \label{eq:GPE3D}
\end{equation}%
where $M$ is the atomic mass, $\bar{V}_{e}=\bar{V}_{\text{trap}}\ +\bar{V}%
_{h}$ is the external potential that consists of two terms, \textit{viz}.,
the transverse confinement and the potential that creates the horizon, $%
\gamma ={4\pi \hbar ^{2}a_{s}}/{M}$ is the coupling constant, $a_{s}$ being
the $s$-wave scattering length, which is supposed to be a function of the
coordinate and time.%
The 3D wave function, $\bar{\Psi}(\bar{\mathbf{r}},\bar{t})$, determines the
number of atoms, $N$, by the normalization condition, $N=\int {|\bar{\Psi}%
|^{2}d^{3}\bar{\mathbf{r}}}$. We fix $N=10^{5}$ of $^{87}$Rb atoms in our
calculations.%

Thus, we consider a thin ring of radius $\bar{R}$ filled by dilute BEC, with
the transverse degrees of freedom frozen by the tight confinement. Assuming
the usual factorized ansatz \cite{jackson1998, pavloff2001,24f,Delgado}, we
set
\begin{equation}
\bar{\Psi}(\bar{\mathbf{r}},\bar{t})=\bar{\psi}(\bar{x},\bar{t})\bar{\phi}(%
\bar{\mathbf{r}}_{\perp },\bar{t}),  \label{factorized}
\end{equation}%
where $\bar{\phi}(\bar{\mathbf{r}}_{\perp })$ is the function of transverse
coordinates (see, e.g. \cite{Gardiner2010}), and the 1D density is
normalized to the total number of atoms:%
\begin{equation}
\int {d^{2}\bar{\mathbf{r}}_{\perp }}|\bar{\phi}|^{2}=1,\quad \bar{n}\equiv |%
\bar{\psi}|^{2}=\int {d^{2}\bar{\mathbf{r}}_{\perp }|\bar{\Psi}|^{2}}.
\label{n}
\end{equation}%
%
%
%
%
Then, averaging Eq. (\ref{eq:GPE3D}) in the transverse plane leads to the
effective 1D GPE (\ref{eq:GPE1D}) with the 1D coupling constant defined as
\begin{equation}
\bar{g}=\frac{4\pi \hbar ^{2}a_{s}}{M}\int {d^{2}\bar{\mathbf{r}}_{\perp }|%
\bar{\phi}|^{4}}.  \label{g}
\end{equation}%

The state of the condensate in the ring with the uniform density is provided
by the plane-wave solution \cite{carusotto2008} of Eq. (\ref{eq:GPE1D}):
\begin{equation}
\bar{\psi}=\sqrt{\bar{n}_{0}} e^{ i(\bar{k}_{0}\bar{x}-\bar{\omega}%
_{0}\bar{t})} ,\quad \hbar \bar{\omega}_{0}=\frac{(\hbar \bar{k}%
_{0})^{2}}{2M}+\bar{V}(\bar{x},\bar{t})+\bar{g}(\bar{x},\bar{t})\bar{n}_{0}.
\label{eq:planeWave}
\end{equation}%
Potential $\bar{V}(\bar{x},\bar{t})$ and nonlinearity coefficient $\bar{g}(%
\bar{x},\bar{t})$ must be mutually matched so as to admit the existence of
the state with uniform density $\bar{n}_{0}$.

Since the wave function must be periodic, $\bar{\psi}(\bar{x})=\bar{\psi}(%
\bar{x}+\bar{L})$, possible values of $\bar{k}_{0}$ and $\bar{k}$ in Eq. (%
\ref{eq:inNoise}) are restricted by the length of the ring: $\left( \bar{k}%
_{0},\bar{k}\right) =2\pi \left( m_{0},m\right) /\bar{L}$, with integers $%
m_{0}$ and $m$ 
determined by the obvious quantization of the velocity circulation:
\begin{equation}
\oint {\bar{\mathbf{v}}d\bar{\mathbf{l}}=2\pi m\hbar /M},
\end{equation}%
where $m$ is the winding number, alias topological charge. 
If we solve Eq. (\ref{eq:GPE1D}) on a grid of
$N_{p}$ points with spacing $\bar{h}_{p}$, then $\bar{k}_{0}=2\pi m/(N_{p}%
\bar{h}_{p})$, where integer $m$ coincides with the winding number.

A desirable spatial distribution of the local sound speeds in regions 1 and
2 in Fig. \ref{model}\ can be produced by making the coupling constant, $%
\bar{g}$ in Eq. (\ref{eq:GPE3D}), spatially and temporally inhomogeneous.
This can be implemented by either tuning the scattering length $a_{s}$ by
means of the Feshbach resonance \cite{carusotto2008,
pitaevskii2016,theocharis2005,theocharis2006,abdullaev2003}, or
longitudinally varying the strength of the transverse confinement \cite{46b}%
. The sound speed in regions 1 and 2 is $\bar{c}_{1,2}=\sqrt{\bar{\mu}%
_{1,2}/M}$, where $\bar{\mu}_{1,2}=\bar{g}_{1,2}\bar{n}$ are the effective
local chemical potentials. For plane wave (\ref{eq:planeWave}) with uniform
density $\bar{n}_{0}$ to be a solution at all values of time, we need to
adjust the axial potential and coupling constant in the two regions as
follows:
\begin{equation}
\bar{V}_{1}(\bar{x},\bar{t})+\bar{g}_{1}(\bar{x},\bar{t})\bar{n}_{0}=\bar{V}%
_{2}(\bar{x},\bar{t})+\bar{g}_{2}(\bar{x},\bar{t})\bar{n}_{0},
\label{eq:uniformCondition}
\end{equation}%
where $-\bar{L}/2 < \bar{x} < \bar{L}/2$.
Actually, a part of the definition of the model is that the combination
written in Eq. (\ref{eq:uniformCondition}) remains constant over the ring.

Finally, to cast the 1D GPE in the normalized form of Eq. (\ref{eq:GPE1D}),
we apply the following rescaling in both regions, and use the notation
without bars: $t=\bar{t}/\bar{\tau}_{1}$, $\omega =\bar{\omega}\bar{\tau}%
_{1} $, $x=\bar{x}/\bar{\xi}_{1}$, $k=\bar{k}\bar{\xi}_{1}$, $\psi =\sqrt{%
\bar{\xi}_{1}}\bar{\psi}$, $g_{1,2}=\bar{g}_{1,2}/(\bar{\xi}_{1}\bar{\mu}%
_{1})$, where $\bar{\tau}_{1}=\hbar /\bar{\mu}_{1}$ and $\bar{\xi}_{1}\equiv
\sqrt{\hbar ^{2}/(M\bar{\mu}_{1})}$ is the healing length in region 1.



The boundary conditions for Eq. (\ref{eq:GPE1D}) being periodic, we used the
split-step fast Fourier transform method \cite{agraval1995} for numerical
simulations. The transition from $g_{1},V_{1}$ to $g_{2},V_{2}$ in region $%
\sigma_x$ is smooth, provided by the potential taken, locally,
as
\begin{equation}
V(x,t)=V_{1}+\Delta V(t)f\left( \frac{t-t_{0}}{\sigma _{t}}\right) f\Big(%
\frac{x-x_{\mathrm{BHH}}}{\sigma _{x}}\Big),  \label{eq:potential}
\end{equation}%
where $-L/2<x<0$, $\Delta V(t)=(V_{2}-V_{1})\theta (t-t_{0})$ [$\theta \left(
t\right) $ is the step function, hence $g$ and $V=V_{1}$ are constant at $%
t\leq 0$], $\sigma _{t}$ is the time of the switch of $V(x,t)$ and $g(x,t)$
from initial to final values, and the switching function is%
\begin{equation}
f(x)=\frac{1}{2}[1+\tanh {(x/2)}].
\end{equation}%
Simultaneously with the variation of
the potential, the nonlinearity coefficient was varied so as to keep $%
V(x,t)+g(x,t)n_{0}$ constant, see Eq. (\ref{eq:uniformCondition}). 

In our simulations we used $x_{\mathrm{BHH}}=-50$ and $t_{0}=5$. We have
checked that the results do not change essentially if potential (\ref%
{eq:potential}) is used without the jump imposed by $\theta (t-t_{0})$,
provided that condition $\sigma _{t}\ll t_{0}$ holds.
The initial condition taken as the uniform condensate with constant $g$ and $%
V=V_{1}$ makes it easier to add the long-wavelength noise to the entire
system or some part of it, if necessary.

Using this setting with variable of $V(x,t)$ and $g(x,t)$ defined as per Eq.
(\ref{eq:potential}) and running hundreds of simulations, it is
straightforward to see that the initial noise added to the uniform
condensate excites all possible long-wavelength modes admitted by nonuniform
$g(x)$ and $V(x)$. We have checked that, for $\sigma _{t}\in \lbrack 0.1,5]$%
, variation of this parameter does not change the dynamics qualitatively,
therefore we here present the results for $\sigma _{t}=0.5$.

Handling the WHH, which always appears in the ring, is more challenging,
therefore we performed simulations for two different switching functions,
the first one was chosen for $0< x < L/2$ as
\begin{equation}
V(x,t)=V_{1}+\Delta V(t)f\Big(\frac{t-t_{0}}{\sigma _{t}}\Big)f\Big(\frac{x_{%
\mathrm{WHH}}-x}{\sigma _{\mathrm{WHH}}}\Big),
\end{equation}%
that provides mirror symmetry for the global potential, under conditions $%
\sigma _{\mathrm{WHH}}=\sigma _{x}$ and $x_{\mathrm{WHH}}=-x_{\mathrm{BHH}}$%
, cf. Eq. (\ref{eq:potential}). One can see that this type of the transition
function cannot allow an arbitrary value of the slope of $c(x)$ at the WHH
point, as it requires an extremely large length of the ring for large $%
\sigma _{\mathrm{WHH}}$ and small $|\partial c/\partial x|\sim (\sigma _{%
\mathrm{WHH}})^{-1}$. To deal with this case, we used, instead of the 
switching scenario defined by Eq. (\ref{eq:potential}), a double-step one, designed as a set of two similar
potentials, see Figs. \ref{SmallLowk} and \ref{SmallHighk}(b) below. These
two functions gradually carry over into each other, allowing one to
manipulate a sufficiently smooth slope near WHH.

\begin{figure*}[tbh]
\includegraphics[width=0.99\textwidth]{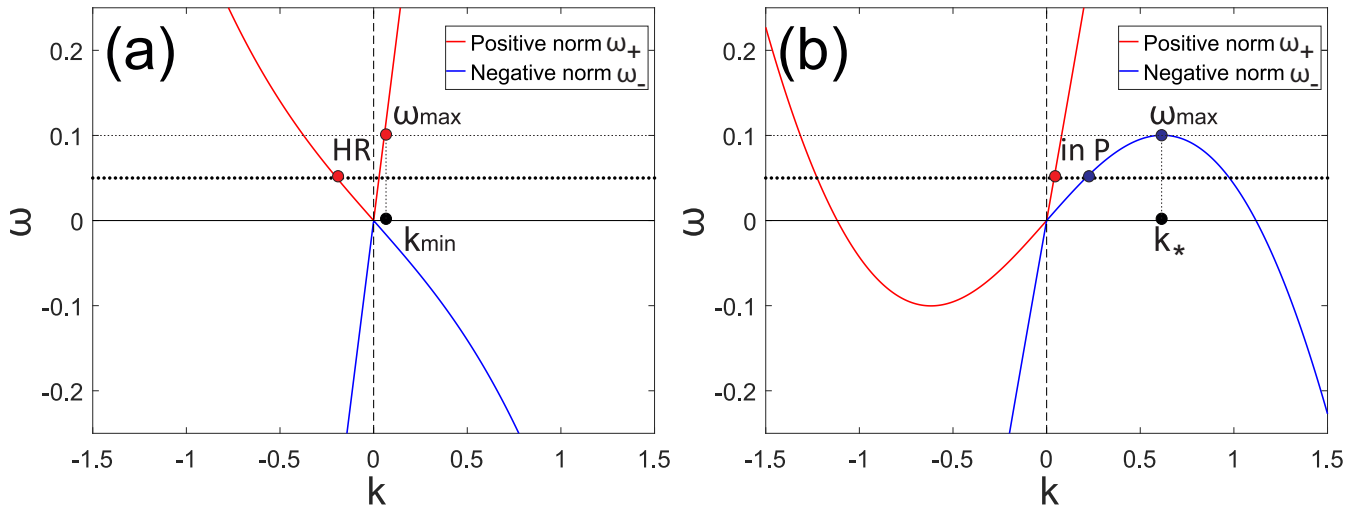}
\caption{ The dispersion relation for regions far from the horizon. The
dispersion curves (a) for $c=c_{1}$ (before BHH) and (b) for $c=c_{2}$
(behind BHH). Panel (a) is relevant for the entire ring, until switching the
horizon on at $t=5$. The HR (Hawking-radiation) mode moves against the flow.
Positive- and negative-norm branches correspond to positive and negative
signs of frequencies in the quiescent condensate. In the condensate with the
flow in the supersonic region, the negative-norm branch acquires positive
values for some values of $k$, which results in the emergence of the $%
\mathrm{P}$ (partner) mode. Modes $\mathrm{P}$ and $\mathrm{in}$ belong to different
dispersion branches and move in the direction of the flow. They represent, respectively, the partner component in the Hawking pair, and a mode which exists even in the absence of the event horizon and moves towards the white-hole horizon. Constant $\protect%
\omega _{\max }$ defines the frequency limit for the analog HR, representing
the maximum frequency on the negative-norm branch in the supersonic region.
Wavenumber $k_{\text{min}}$ represents the restriction on the analog HR
emission for modes with the positive direction of motion. }
\label{dispersion}
\end{figure*}

Parameters of the horizon in the scaled units are similar to those in the
model used in Ref. \cite{carusotto2008} and close to the experimental
parameters of Ref. \cite{steinhauer2016} for the speeds: $c_{1}=1$ ($1$
mm/s), $v=0.74$, $c_{2}=0.5$, $\sigma _{x}=0.5$, $\sigma _{t}=0.5$, with the
corresponding dispersion relations for regions far from the horizon shown in
Fig. \ref{dispersion}. Numerical values of $g_{1,2}$ and $V_{1,2}$ can be
found from relation $\bar{g}=M\bar{c}^{2}/\bar{n}$ and $\bar{V}_{1}=1.6\bar{%
\mu}_{1}$. The initial ring has the length of $\bar{L}=233~\mathrm{\mu }$m
with $\bar{\tau}_{1}=0.73~$ms, $\bar{\xi}_{1}=0.73~\mathrm{\mu }$m, $\bar{c}%
_{1}=1$ mm/s, which is tantamount to $L=320$, $\tau _{1}=1$, $\xi _{1}=1$, $%
\mu _{1}=1$ and $c_{1}=1$ in the scaled units.

\section{Quantum fluctuations and the dispersion relation}

\label{sec:fluctuations_and_dispersion} In real BEC, the presence of quantum
and thermal fluctuations, which are not taken into account by GPE, is
inevitable. It is possible to include these effects by adding Gaussian noise
to the initial wave function, according to TWA. The evolution of the
fluctuations is governed by the Bogoliubov theory \cite{pitaevskii2016}, and the
expectation values of symmetrically ordered observables can be obtained by
taking the stochastic average over the ensemble of evolved wave functions.

As said above, the initial state of the unperturbed system is uniform, and
the potential step is switched on at time $t_{0}>0$, therefore in our
simulations we used the stochastic initial wave function (\ref{eq:inNoise}),
where the summation is performed up to some maximum wavenumber $k$. In the
following section, we discuss an appropriate choice of this limit value, and
consider its impact on the analog HR.

It is straightforward to derive the respective dispersion relation for the
acoustic waves, as small-amplitude collective perturbations running on top
of the uniform-density potential:
\begin{equation}
\omega _{\pm }(k)=vk\pm \sqrt{\frac{k^{2}}{2}\Big(\frac{k^{2}}{2}+2gn_{0}%
\Big)},  \label{eq:dispersion}
\end{equation}%
where wavenumbers $k$ should satisfy the same quantization condition as in
wave function (\ref{eq:planeWave}).
To briefly explain the derivation of Eq. (\ref{eq:dispersion}), we consider
the wave function in a general form,
\begin{equation}
\begin{split}
\psi (x,t)& =e^{-i\omega _{0}t}\Big[\sqrt{n_{0}}e^{ik_{0}x}+\frac{1}{\sqrt{L}%
}\sum_{k\neq 0}\big(\beta _{k}u_{k}(x)e^{-i\omega _{k}t}+ \\
& +\beta _{k}^{\ast }w_{k}^{\ast }(x)e^{i\omega _{k}t}\big)\Big],
\end{split}
\label{eq:generalWavefunc}
\end{equation}%
[cf. the initial condition given by Eq. (\ref{eq:inNoise})], where the
perturbations are assumed to be small in comparison to $\sqrt{n_{0}}$, hence
this expression can be rewritten as $\psi (x,t)=e^{-i\omega _{0}t}\big[\psi
_{0}(x)+\psi ^{\prime }(x,t)\big]$, with $|\psi _{0}(x)|\gg |\psi ^{\prime
}(x,t)|.$ After inserting this ansatz in GPE and linearizing with respects
to $\psi ^{\prime }(x,t)$, one derives a system of the Bogoliubov - de
Gennes equations:
\begin{equation}
\begin{split}
\omega _{k}u_{k}(x)& =(\hat{H}_{0}-\omega _{0}+2gn)u_{k}(x)+g(\psi
_{0})^{2}w_{k}(x), \\
\omega _{k}w_{k}(x)& =(\hat{H}_{0}-\omega _{0}+2gn)w_{k}(x)+g(\psi
_{0}^{\ast })^{2}u_{k}(x),
\end{split}
\label{eq:sys}
\end{equation}%
where $\hat{H}_{0}=-1/2\partial _{xx}^{2}+\hat{V}(x)$. Since $%
u_{k}(x)=u_{0k}e^{i(kx+k_{0}x)}$, $w_{k}(x)=w_{0k}e^{i(kx-k_{0}x)}$ and $%
\omega _{0}=k_{0}^{2}/2+V+gn$, we obtain:

%
\begin{equation}
\begin{split}
(\omega _{k}-k_{0}k-k^{2}/2-gn)u_{0k}-gn_{0}w_{0k}& =0, \\
gn_{0}u_{0k}+(\omega _{k}-k_{0}k+k^{2}/2+gn)w_{0k}& =0.
\end{split}
\label{eq:linearizedSystemSimplified}
\end{equation}%
It is easy to see that dispersion relation (\ref{eq:dispersion}) follows
from the consistency condition of system (\ref{eq:linearizedSystemSimplified}%
), setting $n=n_{0}$ in it. Coefficients $u_{0k},w_{0k}$ ($k\neq 0$) are
solutions of Eq. (\ref{eq:linearizedSystemSimplified}) such that the
corresponding modes $u_{k}(x),w_{k}(x)$ satisfy the normalization condition,
$\int_{0}^{L}\left( {|u_{k}(x)|^{2}-|w_{k}(x)|^{2}}\right) {dx}=1$. Thus,
expressions for ampolitudes of the Bogoliubov modes take the form written in
(\ref{eq:coefs}).


To keep the number of atoms constant, we adjusted density $n_{0}$ in each
simulation. The expression that defines the number of excited atoms in the
uniform condensate was produced in Ref. \cite{ruostekoski2013}. At zero
temperature, it can be written as
\begin{equation*}
N_{s}^{\prime }=\sum_{k\neq 0}\big(|u_{0k}|^{2}+|w_{0k}|^{2}\big)\Big(\beta
_{k}^{\ast }\beta _{k}-\frac{1}{2}\Big)+\sum_{k\neq 0}|w_{0k}|^{2}.
\end{equation*}%
Therefore, the number of atoms remaining in the ground states is $%
N_{c}=N-N_{s}^{\prime }$, and $n_{0}=(N_{c}+1/2)/L.$ Such expressions follow
from the relation between the average over ensemble and quantum average.%
According to Refs. \cite{ruostekoski2013, sinatra2002}, amplitudes $\beta
_{k},\beta _{k}^{\ast }$ of the random perturbation are distributed as
follows:
\begin{equation*}
W(\beta _{k},\beta _{k}^{\ast })=\frac{1}{2\pi \sigma _{\epsilon }^{2}}\exp {%
\Big[-\frac{|\beta _{k}|^{2}}{2\sigma _{\epsilon }^{2}}\Big]},
\end{equation*}%
where $\sigma _{\epsilon }=(4\tanh {(\epsilon _{k}/2k_{B}T_{0}))^{-1/2}}$, $%
T_{0}$ being the temperature of the condensate.

Clear evidence of the analog HR was obtained, both in the zero-temperature
limit and for $T_{0}>0$, in Refs. \cite{carusotto2008,mayoral2011}. Here, we
use the distribution for $T_{0}=0$ and focus on this fundamental case.%

\section{Analog HR (Hawking radiation) near the BHH (black-hole horizon)}

\label{sec:Hawking_radiation} 
To identify the generation of HR from the acoustic BH, we have implemented
the correlation-function method \cite%
{balbinot2008,carusotto2008,steinhauer2014,steinhauer2016,steinhauer2019,recati2009,mayoral20112,fabbri2011}.
It was developed and used in Refs. \cite{balbinot2008,carusotto2008} for the
1D condensate with a continuous transition between two uniform regions. The
simple interpretation of this approach is that the HR and the corresponding
partner particles are correlated with each other, although being located on
opposite sides of the horizon.

We used the normalized density-density correlation function in the form of
\begin{equation}
G(x_{1},x_{2})=\frac{\langle n(x_{1})n(x_{2})\rangle }{\langle
n(x_{1})\rangle \langle n(x_{2})\rangle },  \label{eq:CorrFunc}
\end{equation}%
where averaging is performed over an ensemble of $100$ GPE simulations.
Further increase in the number of simulations does not tangibly affect the
results. All simulations started with the input in the form of the uniform
condensate with quantum noise added to it, as per Eq. (\ref{eq:inNoise}). An
essential role in the analysis is played by the number of modes which are
used in TWA. The number was chosen to satisfy natural restrictions that
allow one to observe signatures of the analog HR. Namely, TWA was proven to
be correct for dilute Bose condensates if the number of modes obeys the
constraint $N>N_{\mathrm{modes}}/2$ Ref. \cite{sinatra2002}. To provide the
presence of $\mathrm{P}$-modes and thermality of the outgoing flux,
frequencies should obey the condition
\begin{equation}
\omega <\omega _{\max },  \label{eq:omegamax}
\end{equation}%
see Fig. \ref{dispersion}, as found in Ref. \cite{macher2009}.%
For a sufficiently low wavenumber, the latter condition also implies that
all modes satisfy the linear dispersion relation, thus maintaining a clear
analogy with HR near a real BH.

Note that inequality (\ref{eq:omegamax}) is written for the inhomogeneous
setting in which the density is uniform, while $V$ and $g$ are not constant
and both horizons are present. In this work, we do not explicitly consider
the eigenstate problem on top of the inhomogeneous background, but rather
use eigenstates produced by the system of the Bogoliubov - de Gennes
equations for the uniform condensates to generate \emph{initial} random
perturbations, as the initial background is uniform, see Eq. (\ref{eq:potential}%
). By running multiple simulations of the ensuing evolution of the
condensate for different realizations of the initial random perturbations,
in the framework of the full GPE, we intend to produce all essential modes
that can be excited after the switch to the inhomogeneous system at $t=t_{0}$.
Accordingly, condition (\ref{eq:omegamax}) was
applied to the initial modes as an estimate for initial frequencies, which
we expect to account for a qualitative change in the behavior of the
correlation function. Therefore, below we refer to $\omega $ of the uniform
setting and the respective dispersion relations displayed in Fig. \ref%
{dispersion}.

The initial noise in Eq. (\ref{eq:inNoise}) involves summation over negative
and positive wavenumbers $k$, but, in our investigation, only positive ones
are relevant, as they correspond to modes moving towards BHH.
Actually, including modes with the negative direction of motion did not
produce any conspicuous change in the numerically computed correlations,
therefore in what follows below the initial conditions include only lower
positive-$k$ modes.

\begin{figure}[tbh]
\includegraphics[width=3.4in]{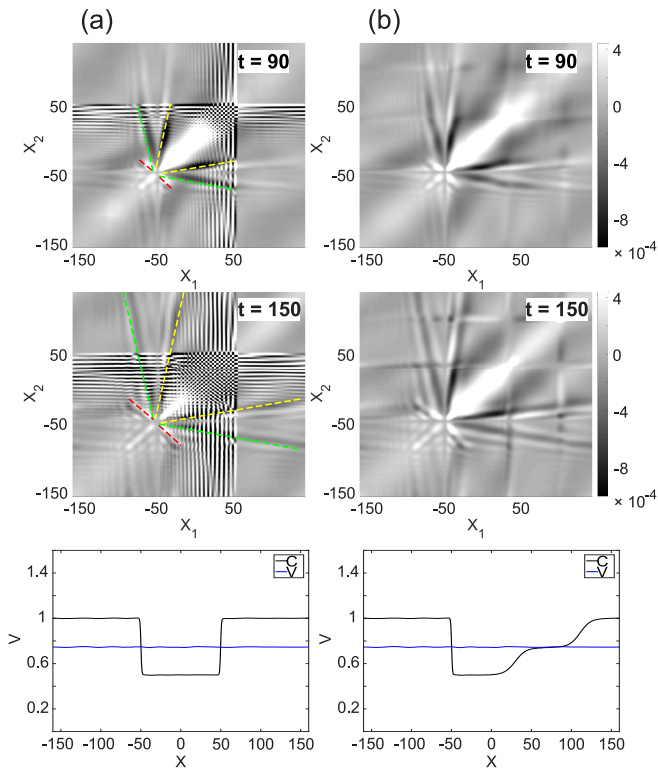}

\caption{Snapshots of the correlation function $n\protect\xi _{1}\cdot
(G(x_{1},x_{2})-1)$ for $t=90$ and $t=150$ for two different profiles of the
local speed of sound with three initial modes and the ring's length $L=320$.
The bottom row represents the distribution of the velocity (blue line) and local
speed of sound (black curve) along the ring. (a) Symmetric BHH and WHH
(black- and while-hole horizons) with sharp gradients of the local speed of
sound near both horizons. (b) Speed of sound gradient is smoothed near the
WHH. Note that the checkerboard pattern in the supersonic region, which is
clearly seen in (a), vanishes in (b), when the local speed of sound
gradually increases near the white horizon. Yellow, green, and red lines
show expected positions of $\mathrm{P-in}$\textrm{, }$\mathrm{HR-in}$\textrm{%
, }$\mathrm{HR-P}$ correlation tongues. }
\label{SmallLowk}
\end{figure}

First, we address the symmetric potential with initial parameters given
above ($L=320$, $x_{\mathrm{BHH}}=-50$, $x_{\mathrm{WHH}}=50$, $\sigma
_{x}=\sigma _{\mathrm{WHH}}=0.5$). Accordingly, in condition $\omega <\omega
_{\max }$ we set $\omega _{\max }\ \approx 0.1$. In this case, the system
admits three modes with the initial frequency below $\omega _{\max }$.
The obtained results for the
correlation function are displayed in Fig. \ref{SmallLowk}. Colored lines in
the figure correspond to the expected correlations between different
particle pairs belonging to the dispersion relation in Fig. \ref{dispersion}%
. The slope of the colored segment relative to the $x$-axis is determined by
expressions for the sound speed of the Bogoliubov excitations:
\begin{equation*}
\tan (\theta _{y})=\frac{v-c_{2}}{v+c_{2}},\tan (\theta _{g})=\frac{v-c_{1}}{%
v+c_{2}},\tan (\theta _{r})=\frac{v-c_{1}}{v-c_{2}},
\end{equation*}%
where $\theta _{y},\theta _{g}$, and $\theta _{r}$ refer to the yellow,
green and red lines, respectively, in Fig. \ref{SmallLowk}. The length of
each segment represents the expected length of the correlation tongues at
the corresponding times, where \textquotedblleft tongues" designate the
line-shaped correlation regions that have one end located near BHH and other
end growing in time in a certain direction.%
It is seen that, in spite of the quantization of $k$, all the correlation
tongues are visible for both potentials and they agree well with the
predictions based on the dispersion relations. In Fig. \ref{SmallLowk} it is
seen that the correlation pattern alters at later times [in Fig. \ref%
{SmallLowk}(a) at $t\simeq 150$] because a checkerboard pattern appears as a
result of the existence of a cavity between the horizons \cite%
{steinhauer2014,mayoral2011}. To resolve this problem, we decreased the
value of $|\partial c/\partial x|\sim \left( \sigma _{\mathrm{WHH}}\right)
^{-1}$, which is responsible for the mixing of modes at the horizon. Applying
the double-step potential, we could make $|\partial c/\partial x|$ small enough, and thus avoid the destructive impact of such effects, by reducing the
checkerboard correlations to just two crosses created by the scattering on
each step. While we observe the correlations in both patterns for the
current value of the ring's radius, the use of such decaying $|\partial
c/\partial x|$ is necessary for smaller radii that are considered below.

\begin{figure}[htb]
\includegraphics[width=3.3in]{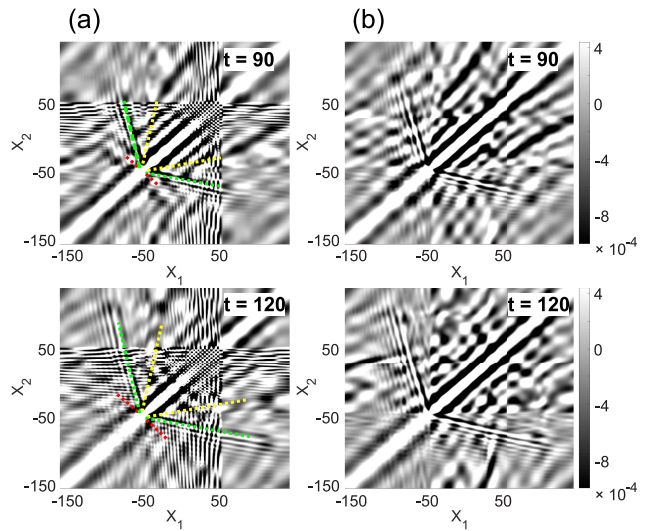}

\caption{The same as in Fig. \protect\ref{SmallLowk} (without the bottom
panels) for $L=320$, but inputs with ten initial modes. Note that the
checkerboard pattern vanishes for the smoothed white horizon, as in Fig.
\protect\ref{SmallLowk}. }
\label{SmallHighk}
\end{figure}

The next step is to apply the initial noise, which contains modes that are
located at $\omega >\omega _{\max }$, in terms of the dispersion relation.
The corresponding correlations, produced by ten lowest modes, are shown in
Fig. \ref{SmallHighk}. As above, we see the same correlations for steep and
smoothed slopes at WHH, with a qualitative difference in the checkerboard
correlation patterns. Moreover, only $\mathrm{HR-in}$ correlation tongues
(i.e., ones between excitation modes of the $\mathrm{HR}$ and $\mathrm{in}$
types) are present, and there is no evidence of $\mathrm{P-in}$ or $\mathrm{%
HR-P}$ tongues. We also note that obtained correlation tongues are expanding
a bit faster than expected, which may be a consequence of being beyond the
linear-dispersion regime.

\begin{figure}[tbh]
\includegraphics[width=3.4in]{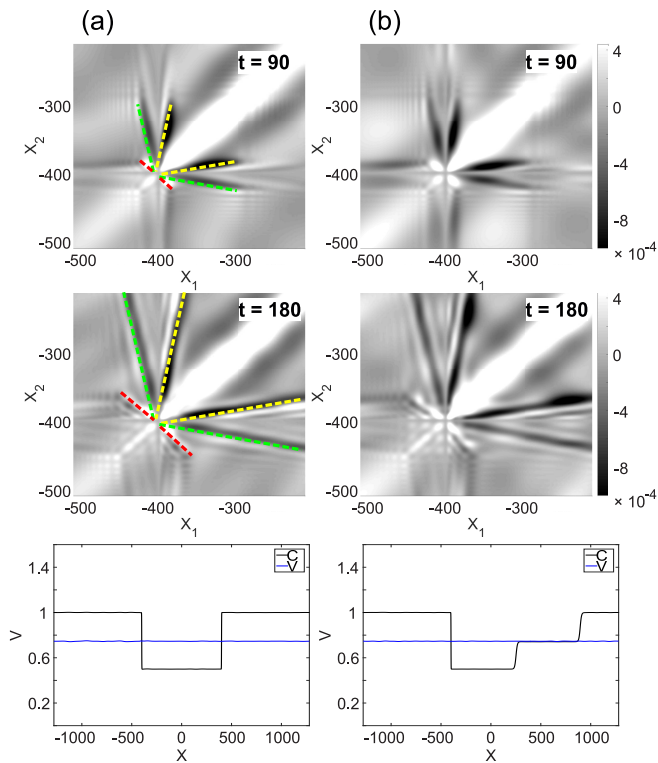}
\caption{Correlation function $n\protect\xi _{1}\cdot (G(x_{1},x_{2})-1)$
for steep (a) and smoothed (b) slopes for 23 initial modes and $L=2560$. The
bottom row shows velocity (blue line) and local speed of sound (black line)
profiles near WHH. The spatial scale in the ($x_{1},x_{2}$) plane is the
same as in Fig. \protect\ref{SmallLowk}.}
\label{LargeSmallk}
\end{figure}

We have performed similar simulations for the eight times enlarged region
and distance between the horizons ($L=2560$ or $\approx 1.87$ mm, in
physical units), keeping the same parameters of both horizons as above. The
larger area makes it possible to admits more initial modes that lie below $%
\omega _{\max }$. We have thus performed the simulations for $23$ lowest
modes belonging to the linear-dispersion regime. 
In Fig. \ref{LargeSmallk} one can see the entire range of the expected
correlations, and the expected positions of the tongues very well coinciding
with the highlighted lines in the correlation pattern. These correlations
feature relatively high intensity, which is about $4\times 10^{-4}$ for the $%
\mathrm{HR-P}$\textrm{\ }tongues. This value is still $\simeq 10$ times
smaller than that reported in Ref. \cite{carusotto2008}, which is explained
by our choice of the diluteness parameter, $\xi _{1}n=1/g_{1}=312.5\gg 1$,
that defines the intensity of the correlation signal. No drastic difference
is seen between the correlations near BHH for both potential slopes at WHH.
This fact is explained by a finite speed of excitations and relatively large
distance between the two horizons.
\begin{figure}[tbh]
\includegraphics[width=3.4in]{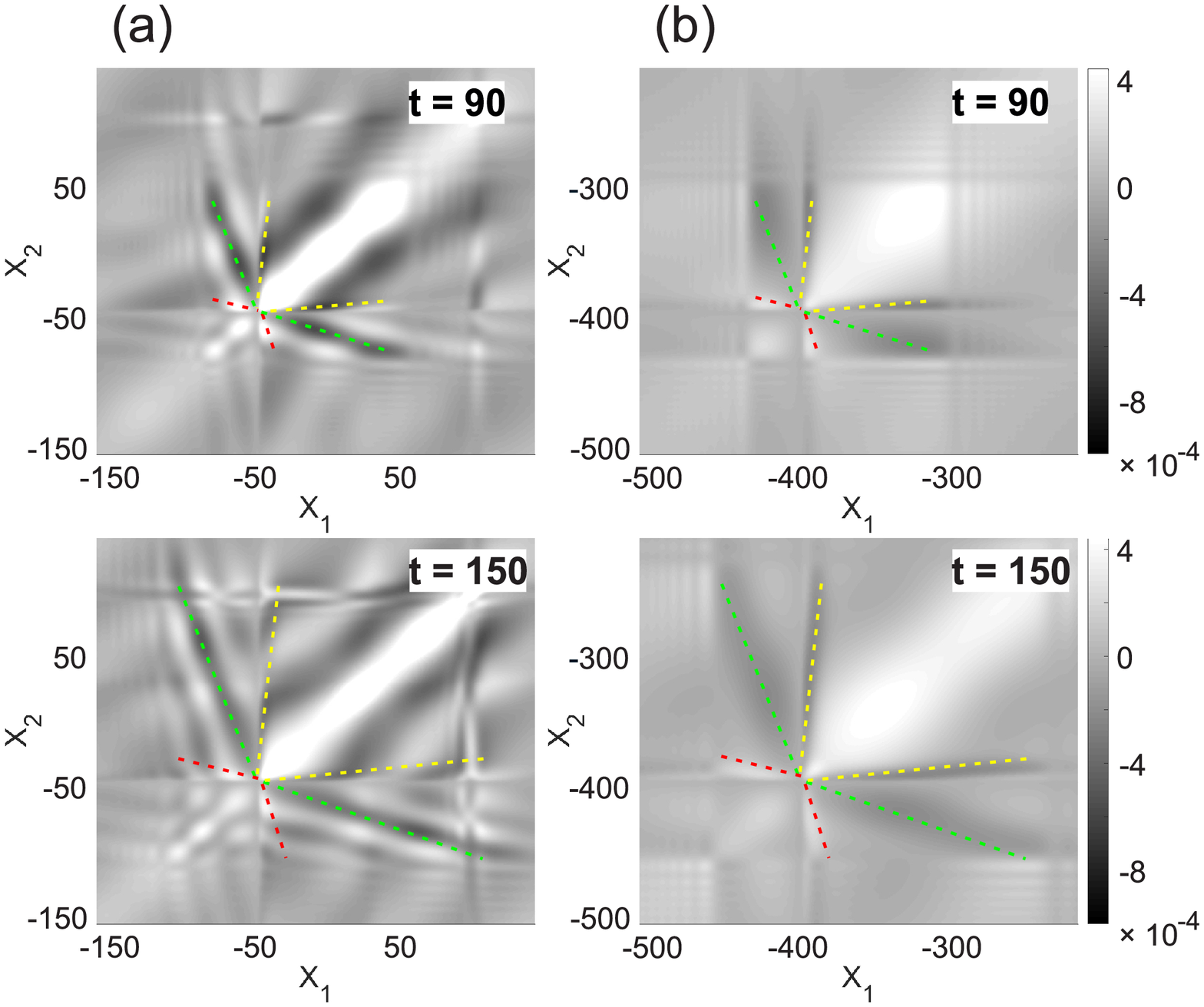}
\caption{Vanishing of $HR-P$ correlations for the ring's length $L<L_{%
\mathrm{cr}}$ for a ring with $v=0.61$ $(m=31)$ and values of $|\partial
c/\partial x|$ much smaller near WHH, in comparison to BHH. Column (a)
pertains to the ring with $3$ lowest modes and $L=320$, where even the
lowest $k$-mode exceeds the frequency limit, and (b) provides snapshots of
weak correlations for a larger ring ($L=2560$) and $10$ lowest initial
modes. The scale of the correlations is the same in all snapshots. }
\label{CriticalSize}
\end{figure}

The results obtained for the initial noise containing modes with $\omega
<\omega _{\max }$ in a large ring ($L=2560$) share main properties with the
previous simulations: there is no difference between correlation patterns
near BHH for two different slopes at WHH (at least in the course of the
simulation time), and the correlation pattern, as one may expect,
features solely $\mathrm{HR-in}$ correlation tongues. The same happens with
modes that have energy above $\omega _{\max }$.

The inference is that the possibility to observe the analog HR correlations
(for fixed parameters of the flow and horizon) depends on the number of
modes that have their frequencies below and above $\omega _{\max }$, as
predicted by dispersion relation (\ref{eq:dispersion}) for the uniform
density. Furthermore, the effect is strongly affected by the size of the
ring. The former aspect reflects the fact that, to produce the HR effect,
one needs to have particles belonging to the frequency from the
negative-norm branch of the dispersion relation in the laboratory reference
frame. This criterion was also discovered analytically for the elongated 1D
condensate with the single horizon \cite{macher2009}, and it is true for our
system as well, even if we apply it to the initial noise. The latter aspect
implies that, by enlarging the length of the ring, one increases the number
of allowed modes, which, in the limit of the ring with an infinite radius
leads to the continuous frequency spectrum and an infinite number of modes at $%
\omega <\omega _{\max }$ past $t=t_{0}$. On the other hand, when the length
of the ring decreases, the discreteness of the system's spectrum makes a
great difference in the predicted observations: while some modes stay within
the linear-dispersion region in a large-radius torus, the same set of the
modes can partially or even completely exceed the frequency limit for a
small ring. This change is accompanied by the transition from three pairs of
the correlation tongues to the single $\mathrm{HR-in}$ surviving one, or
even the absence of any correlations, if a sufficiently large number of
modes are located at $\omega >\omega _{\max }$ ).
Lastly, it is expected that, in an elliptically deformed torus, the
dependence of the HR effect on $L$ remains the same, in the first
approximation. 

Another important issue concerning our system with two horizons is its
stability, which is sensitive to boundary conditions and the presence of the
different horizons \cite{Barcelo2006}. For the initial noise that does not
strongly violate the frequency condition, we have found the steep horizon to
be subject to eventual instability, with the respective lifetime $t_{\mathrm{%
decay}}\simeq 600$ ($\approx 0.438$ s, in physical units) for $L=320$. At
larger times, simulations demonstrate that the development of the
instability near WHH produces dark-soliton-like structures, and, finally, it
leads to decay of both horizons. While similar behavior was reported in Ref.
\cite{garay2001}, in the present system instabilities grow near WHH, rather
than BHH. On the other hand, the system with a smoothed horizon shows no
instability (at least, in the course of long-time simulations for $t\leq 1$
s). The configurations in larger rings, with a bigger distance between the
horizons, tend to be more stable than in smaller ones.

Due to the above-mentioned fact that the presence of the $\mathrm{HR-P}$
correlations tongue is determined by the number of initial modes below and
above the critical value of $\omega _{\max }$, the value of the lowest
wavenumber is restricted by the length of the ring. Therefore, there exists
a critical size making it impossible to observe the HR correlations, as no
Bogoliubov mode may satisfy the constraint (roughly speaking, this
conclusion resembles the known property of the modulational instability,
which is suppressed by periodic boundary conditions if the ring's length
falls below the respective critical value \cite{MI}).%

It is worth noting that restrictions on the size of the ring-shaped
condensate have already been discussed in work \cite{garay2001}. That
constraint originates from the phononic regime for the perturbations and
sufficient conditions for the existence of the sonic horizon, giving
critical value $R_{\mathrm{cr}}^{\prime }=\sqrt{2\pi /Ng}$ or $R_{1\mathrm{cr%
}}^{\prime }(g_{1})=0.14$, $R_{2\mathrm{cr}}^{\prime }(g_{2})=0.28$, which
are significantly smaller than the critical radii obtained above. Moreover, $%
R_{\mathrm{cr}}^{\prime }$ depends on the particle number, which was kept
constant in Ref. \cite{garay2001}. However, we change $N$ proportionally to
the $L$ in order to keep $n_{0}$ and $c_{1,2}=\sqrt{g_{1,2}n_{0}}$ constant.
Consequently, our estimate for $R_{\mathrm{cr}}$ depends solely on
parameters of the horizon, $v,c_{1,2}$. The restriction on the radius is
determined by the presence or absence of correlations as a signature of the
analog-HR effect. The critical size $R_{\mathrm{cr}}$ for our system can be
evaluated from the condition that the frequency, which corresponds to the
lowest wavenumber pursuant to Eq. (\ref{eq:dispersion}), is equal to $\omega
_{\max }=\max [\omega _{-}(k;g_{2})]$, see Fig. \ref{dispersion}. The
respective relation which determines the critical radius, $R_{\mathrm{cr}%
}=1/k_{\mathrm{cr}}$, becomes

\begin{equation}
\omega _{\max }=\frac{v}{R_{\mathrm{cr}}}+\sqrt{\frac{1}{2R_{\mathrm{cr}}^{2}%
}\Big(\frac{1}{2R_{\mathrm{cr}}^{2}}+2g_{1}n_{0}\Big)}.
\end{equation}%
For $1/R_{\mathrm{cr}}\ll 1$ it simplifies to
\begin{equation}
R_{\mathrm{cr}}=\frac{v+c_{1}}{\omega _{\max }}=\frac{2v(v+c_{1})}{2k_{\ast
}(v^{2}-c_{2}^{2})-k_{\ast }^{3}},  \label{eq:crit}
\end{equation}%
where
\begin{equation*}
k_{\ast }\equiv c_{2}\sqrt{-\Big(2-\frac{v^{2}}{2c_{2}^{2}}\Big)+\frac{v}{%
c_{2}}\sqrt{2+\frac{v^{2}}{4c_{2}^{2}}}}.
\end{equation*}

It is relevant to compare the observed correlations for different critical
radii evaluated from Eq. (\ref{eq:crit}). With our initial parameters of
BHH, which allowed the observation of the $\mathrm{HR-P}$ correlations, the
critical length of the ring is $L_{\mathrm{cr}}\ \approx 110$, that
corresponds to critical radius $R_{\mathrm{cr}}\ \approx 17$ (or $\bar{R}_{%
\mathrm{cr}}\approx 13$ $\mathrm{\mu }$m, in physical units). It is three
times smaller than the radius of the ring in the simulations, $R\approx 51$.
However, it remains a challenging objective to produce evidence of the
disappearance of the $\mathrm{HR-P}$ correlations, as the noise with the
amplitude used in the above simulations suppresses all correlations at $%
\bar{R}\simeq 13$ $\mathrm{\mu }$m.
Therefore, we have changed parameters of the horizon to $v=0.61$ and $L_{%
\mathrm{cr}}=355$, so as to make the critical size of the ring slightly
larger than $L=320$ (which is good to observe correlations), and conducted
the simulations for a smoothed slope at WHH. In Fig. \ref{CriticalSize} we
observe the absence of $\mathrm{HR-P}$ correlations for the three lowest
modes. On the other hand, simulations produced visible $\mathrm{P}$\textrm{$%
-in$} correlations (and negligible $\mathrm{HR-P}$ ones) for the ten lowest
modes. Thus, we have performed numerical simulations of the nonuniform (at $%
t>t_{0}$) condensate and compared the results for the rings with the radius
taken above and below the critical value $R_{\mathrm{cr}}$.

We stress that our consideration of the HR effect is based mostly on the
direct numerical analysis of the density-density correlations. Surprisingly,
predictions based on the oversimplified estimates of $R_{\mathrm{cr}}$ 
found for a
uniform condensate appears to be in good agreement with direct numerical
simulations. 
\section{Conclusions} \label{sec:conclusions} 
We have investigated the possibility to generate acoustic HR\ (Hawking
radiation) in the superfluid ring-shaped BEC. For this purpose, we have
introduced the double-step potential that minimizes the emulated Hawking
temperature near the WHH (white hole horizon), where instabilities may
occur. The desirable region with the supersonic flow and uniform density
distribution of the condensate may be designed using the
spatiotemporally-modulated interaction constant, adjusted to the selected
potential. These features make the system considered here sufficiently
stable and convenient for the analysis of the acoustic analog of HR in the
rings.

We have addressed basic properties of the analog HR in the ring-shaped
matter-wave configurations with different radii. The HR is quantified by the
location and shape of tongue-shaped correlation patterns, which, in turn,
are well predicted by the dispersion relation for the quasi-uniform
condensate in the limit of low wavenumbers. To this end, small random
perturbations are added to the uniform system, in the framework of TWA
(truncated Wigner approximation), and then multiple simulations of the full
GPE are run for different realizations of the initial random perturbations,
while the setting switches from spatially uniform to the nonuniform one. The
so produced results for the correlation patterns are stable against
variations of the spatiotemporal modulation of the potential and coupling
constant.

An important circumstance is that the patterns are sensitive to the number
of initial modes admitted by the ring, depending on its radius, below and above the
frequency limit, $\omega _{\mathrm{\max }}$. Varying these parameters, we
have observed different tongue-shaped correlation patterns: from three pairs
of tongues to a single one for a sufficiently large number of modes. The
discreteness of the frequency and momentum spectra in the ring produces a
dramatic effect on the properties of HR. Below the critical radius of the ring,
even the lowest mode has its frequency above the $\omega _{\mathrm{\max }}$,
so that $\mathrm{HR-P}$ correlations disappear. Thus, no acoustic HR takes
place in the ring-shaped superflows with the radius falling below the
critical value. Remarkably, a rough estimate for the critical radii, given by Eq. (\ref{eq:crit}), which is obtained from the dispersion relation for the \emph{homogeneous} condensate agrees with results of direct numerical simulations of the perturbed dynamics of the inhomogeneous condensate.

It is relevant to compare the minimum radius of the ring, necessary for the
generation of HR, with the situation for real (astrophysical) BHs (black
holes). While they lose their mass through HR extremely slowly, the Hawking
temperature dramatically increases for small BHs with mass $M$: $T_{H}=\hbar
c^{3}/(8\pi Gk_{B}M)=6.169\times 10^{-8}(M_{\odot }/M)$ K. In this
connection, it is relevant to mention that quantum effects are believed to
be crucially important for BHs with the Planck-scale mass, $M\sim m_{P}=%
\sqrt{\hbar c/G}$. A well-known puzzle in the quantum theory of gravity is
the final fate of such BHs. There are good grounds to assume \cite%
{macgibbon1987} that HR is suppressed for sufficiently small BHs when their
size, $r_{g}=2GM/c^{2}$, becomes comparable to the Compton wavelength, $%
\lambda _{C}=h/(m_{P}c)$, associated with the BH of the Planck's mass. On the other hand, we have found that there is also a critical size of the system, related to quantum effects in the toroidal geometry. The critical size depends on the spatial structure of the superflow near the BH horizon, which determines the strength of the analog "surface gravity". Certainly, the properties caused by the quantization, i.e., the discreteness of the frequency and momentum spectra, become essential when the size of the ring shrinks at a fixed value of the slope of the speed of sound. When the system's radius falls below the critical value, $R_{\mathrm{cr}}$, the analog HR disappears. Accordingly, in terms of real BHs, small
non-radiating primordial black holes, which are conjectured to be created in
great numbers in the early universe, might survive and become an ingredient
of dark matter \cite%
{carr2016,2020arXiv200212778C,2019JCAP...10..046L,pacheto2020}.

\section{Acknowledgment}
The work of B.A.M. is supported, in part, by the Israel Science Foundation
through grant No. 1286/17.


\bibliography{references}

\end{document}